\documentclass[manuscript]{aastex}
\usepackage{emulateapj5,timesfonts,graphicx,epsfig,amsbsy}



\def\***#1{{\sc #1}}
\def\plan#1{\relax}
\def\Plan#1{\relax}
\def\PLAN#1{\relax}

\renewcommand{\arraystretch}{1.2}
\def\etal{et al.}
\def \be {\begin{equation}} 
\def \en {\end{equation}} 
\def \eo {{g}_1} 
\def \et {{g}_2}

\def\eg{{\it e.g.}}  

\def\spose#1{\hbox to 0pt{#1\hss}} 
\def\simlt{\mathrel{\spose{\lower 3pt\hbox{$\mathchar"218$}}%
    \raise 2.0pt\hbox{$\mathchar"13C$}}}
\def\simgt{\mathrel{\spose{\lower 3pt\hbox{$\mathchar"218$}}%
     \raise 2.0pt\hbox{$\mathchar"13E$}}}
\newcommand\nabl{\boldsymbol{\nabla}}
\newcommand\bchi{\boldsymbol{\chi}}

\begin{document}
\shorttitle{Lens Mapping Algorithm} \shortauthors{Saini and
Raychaudhury}

\title{A Lens Mapping Algorithm for Weak Lensing}

\author{Tarun Deep Saini$^1$ and Somak Raychaudhury$^{2,1}$}
\affil{$^1$Inter-University Center for Astronomy \& Astrophysics, Pun\'e
411 007, India}
\affil{$^2$School of Physics \& Astronomy, University of Birmingham,
Birmingham, United Kingdom}

\email{saini@iucaa.ernet.in; somak@star.sr.bham.ac.uk}

\begin{abstract}

We develop an algorithm for the reconstruction of the two-dimensional
mass distribution of a gravitational lens from the observable
distortion of background galaxies. From the measured reduced shear
$\gamma_i/(1-\kappa)$ the lens mapping is obtained, from which a mass
distribution is derived. This is unlike other methods where the
convergence $\kappa$ is directly obtained.  We show that this method
works best for sub-critical lenses, but can be applied to a critical
lens away from the critical lines.  For finite fields the usual
mass-sheet degeneracy is shown to exist in this method as well.  We
show that the algorithm reproduces the mass distribution within
acceptable limits when applied to simulated noisy data.

\end{abstract}


\section{Introduction}\label{sec:intro}

The mapping of the distribution of matter in extended regions around
rich clusters of galaxies, from the systematic distortions of
background galaxies due to gravitational lensing, has increasingly
become feasible and popular over the last decade.  This \emph{shear}
is measured from the quadrupole moments of the images of the
background galaxies in terms of the local `polarization' of an image
compared to its assumed intrinsic circular form.  In order to remove
the effect of the source ellipticity distribution, the parameters are
averaged over a large number of source galaxies.  The
two-dimensional mass distribution of the lens is reconstructed from
the shear map away from the critical lines (`weak
reconstruction'). Following the pioneering work of Tyson \etal\ (1990),
the idea was quantitatively developed by Kaiser and Squires (1993,
also Kaiser 1995 and Squires and Kaiser 1996) and Seitz and Schneider
(1995, 1996). Since then, several interesting variants on this theme
have appeared in the literature, \eg\ using maximum likelihood cluster
reconstruction (Bartelmann \etal\ 1996, Seitz \etal\ 1998), or using
methods based on the variational principle (Lombardi and Bertin 1999)
or maximum entropy (Bridle \etal\ 1998).

Since the shear data contain complete information about the mass
distribution in two independent fields $\gamma_1 /(1-\kappa)$ and
$\gamma_2/(1-\kappa)$, it is possible to obtain several algorithms to
estimate the mass distribution from the measured shear.  The observed
shear is available only on a finite grid. Due to the unknown intrinsic
distribution of the ellipticities of the source galaxies, and the effect
of the distortion of the PSF due to observing conditions (seeing,
tracking etc.), the measured shear is noisy.  It is therefore desirable to
develop new algorithms in the hope that they might be able to deal with
the noise better than other methods.

In this paper we develop an algorithm (LM: Lens Mapping algorithm) for
the lens mass reconstruction from the measured reduced shear, $g_i =
\gamma_i/(1-\kappa)$. The method involves two steps. First, the  lens
mapping is reconstructed from the reduced shear, which can be done
uniquely with the assumption that the lens mapping goes to
identity far away from the lens.  As the second step, we show in
\S\ref{sec:algorithm} that for a sub-critical lens, the surface mass
density can be reconstructed completely from the derived lens mapping.
In reality, however, the measured shear is available only in a finite
region of the lens plane. In this case the lens mapping cannot be
uniquely obtained and hence the LM algorithm exhibits a mass-sheet
degeneracy, which is also present in other methods of mass
reconstruction.  In \S\ref{sec:finite}, we characterize this
degeneracy in the context of the LM algorithm. The performance of this
method with discretely sampled, noisy data is dealt with in
\S\ref{sec:noisy}, where we demonstrate the various features of the LM
algorithm, by reconstructing the mass distribution for an analytically
given shear field, and separately showing the effects of discrete
sampling and of uniform noise (due to measurement error) on the input
data.

\begin{figure*}
  \newlength{\figwidth}
  \setlength{\figwidth}{\textwidth}
  \addtolength{\figwidth}{-\columnsep}
  \setlength{\figwidth}{0.5\figwidth}
  
  \begin{minipage}[t]{\figwidth}
    \mbox{}
    \vskip -7pt
    \centerline{\includegraphics[width=0.82\linewidth,angle=0]{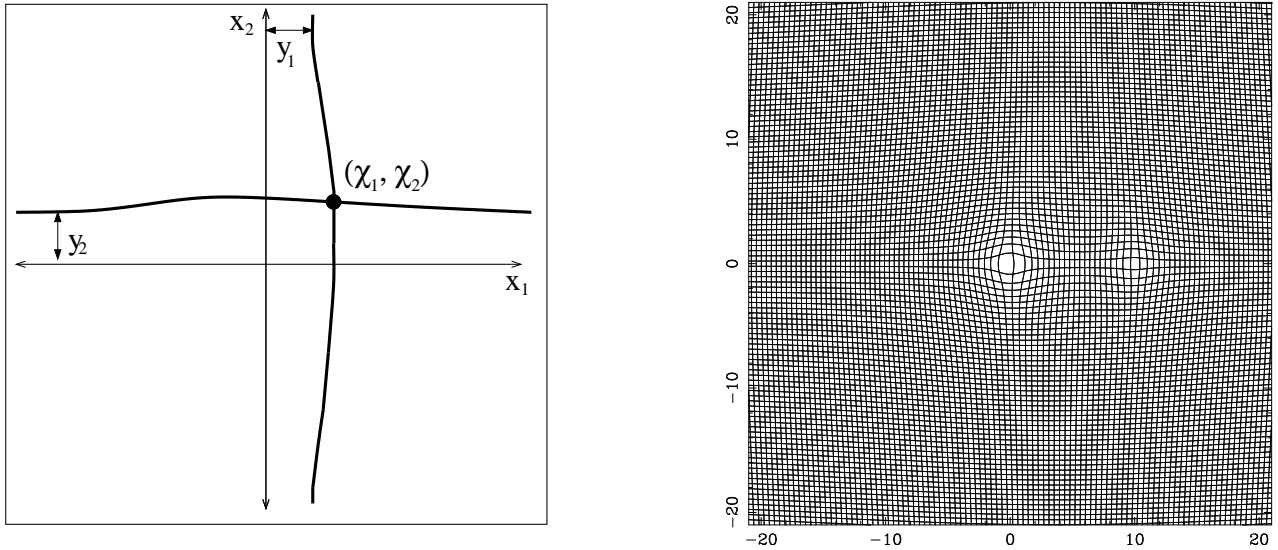}}
  \end{minipage}
  \hfill
  \begin{minipage}[t]{\figwidth}
    \mbox{}
    \vskip -7pt
    \centerline{\includegraphics[width=0.82\linewidth,angle=0]{grid.ps}}
  \end{minipage}
    \caption{
[a] The coordinate system used in
this paper.  We show the lens-plane images of two straight lines
passing through ($y_1$, $y_2$) in the source plane; [b] An
equally-spaced Cartesian grid (in the source plane) is lensed by a
two-component mass model, both PICMDs with zero ellipticity and $k_0 =
0.8$ and $0.5$ respectively (described in \S\ref{sec:algorithm}). 
The axes are marked in units of the core radius $r_c$, assumed to be
same for both the models. The separation between the centers of the
two distributions is $10\,r_c$.  \label{fig:grid} }

\end{figure*}

\section{Weak Lensing and the Mass-Sheet Degeneracy }

For most practical applications of gravitational lensing the lens can
be considered to be thin. Under the small angle approximation the lens
equation is given by
\begin{equation}
{\mathbf y} = {\mathbf x} - \nabl \psi({\mathbf x})\,\,,
\label{eq:lens}
\end{equation}
where the source angular position is denoted as ${\mathbf y}$, the
image angular position as ${\bf x}$ and the relativistic potential
$\psi$ satisfies the equation $\kappa\, ({\bf x})\equiv
\Sigma/\Sigma_{\mathrm{crit}} =\frac{1}{2} \nabl ^2 \psi$, where
$\Sigma_{\mathrm{crit}} =(c^2/4 \pi G)(d_{\rm s}/d_{\rm l} d_{\rm
ls})$. Here $ d_{\rm s},d_{\rm l}$ and $d_{\rm ls} $ are
respectively the angular diameter distance to the source, to the lens
and between the lens and the source. The coordinates for source and
image are small compared to unity and we can treat their components as
Cartesian. It is convenient to define two planes: that containing the
source is called the source plane and that containing the cluster,
the lens plane (or image plane).

The coordinate differentials in the source plane and the corresponding
differentials 
in the lens plane are related through $dy_i = M^{-1}_{ij} dx_j$ and the
inverse of the magnification matrix $M$ is given by
\begin{equation}
M^{-1} = (1-\kappa)
\left( \begin{array}{cc} 1-g_1 & - g_2 \\ -g_2 & 1 + g_1 \\
\end{array} \right ) \>,
\label{eq:matobs}
\end{equation}

where ${\mathbf g} \equiv g_1 + ig_2 = (\gamma_1
+i\gamma_2)/(1-\kappa)$, $\gamma_1= \frac{1}{2} ( \psi_{,11} -
\psi_{,22})$ and $\gamma_2 =\psi_{,12}$ and the subscripts denote
differentiation with respect to the two components of the image
coordinates. From the quadrupole moments of the surface brightness of
the images of the background galaxies the reduced shear can be
measured unambiguously in the regions where $\kappa < 1$ (for
observational details see Kaiser 1999 or Bartelmann \& Schneider
2000). 

The continuity of ${\mathbf y} (x_1,x_2)$ implies 
$\partial^2_{jk}y_i= \partial^2_{kj} y_i$, which, along with,
eq.~(\ref{eq:matobs}) gives
\begin{equation}
{\rm K}_{,l} {\rm G}_{ij}^{-1}- {\rm K}_{,j} {\rm G}_{il,j}^{-1} =
{\rm G}_{il,j}^{-1} - {\rm G}_{ij,l}^{-1}\,\,,
\label{eq:tensor}
\end{equation}
where $\rm{K}\equiv \ln (1-\kappa)$ and $G^{-1} \equiv M^{-1}/(1-\kappa)$.
On multiplying (\ref{eq:tensor}) by the inverse of ${\rm
G}^{-1}$, and taking the trace of the resulting equation gives us
\begin{equation}
{\nabla}_l \ln (1-\kappa)  = \sum_{i=1}^{2} \sum_{j=1}^{2} \left [ {\rm G}_{ij}
({\rm G}_{il,j}^{-1} - {\rm G}_{ij,l}^{-1}) \right]\>.
\label{eq:kaiser}
\end{equation}
This equation was first derived by Kaiser (1995).  It is clear that
replacing $1-\kappa$ on the left hand side with $\lambda(1-\kappa)$,
where $\lambda$ is a constant, does not affect the equation.
Therefore any particular solution of this equation can be used to
obtain a one-parameter degenerate family of functions, all of which
satisfy (\ref{eq:kaiser}). This is known as the mass-sheet degeneracy.

\begin{figure*}
  \setlength{\figwidth}{\textwidth}
  \addtolength{\figwidth}{-\columnsep}
  \setlength{\figwidth}{0.5\figwidth}
  
  \begin{minipage}[t]{\figwidth}
    \mbox{}
    \centerline{\includegraphics[width=0.8\linewidth,angle=0]{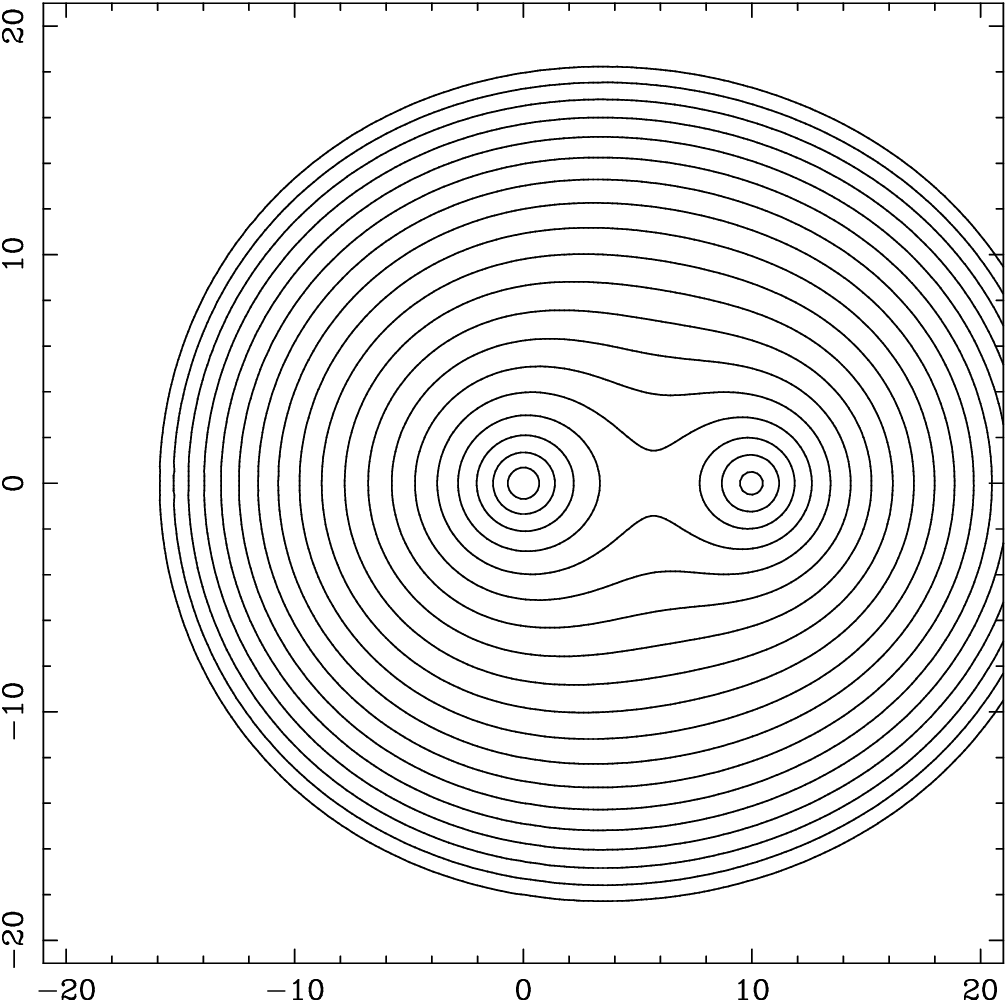}}
\caption{The reconstructed mass distribution, using the Lens Mapping
Algorithm, for the model used in Fig.~\ref{fig:grid}(b), shown as
contours of equal surface density $\kappa$.  The axes are marked in
units of the core radius $r_c$.  \label{fig:contours}}
  \end{minipage}
  \hfill
  \begin{minipage}[t]{\figwidth}
    \mbox{}
    \centerline{\includegraphics[width=0.8\linewidth,angle=0]{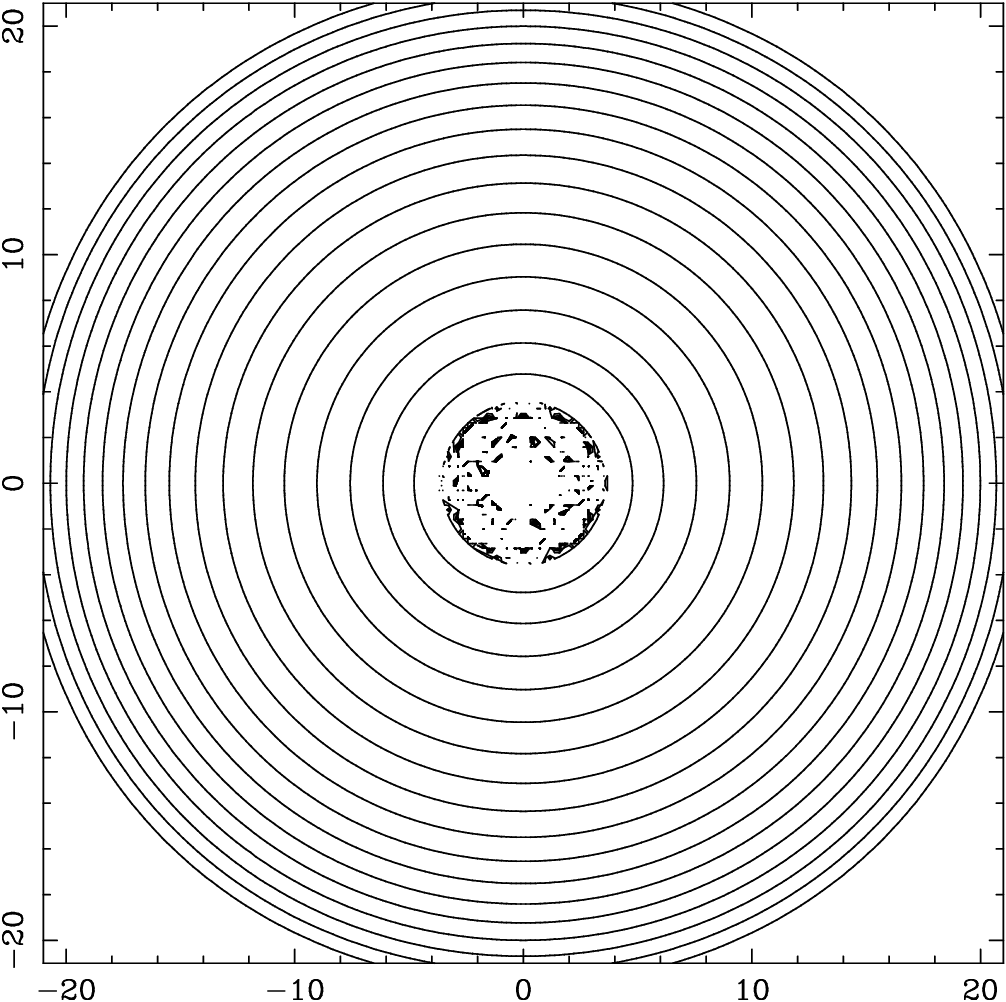}}
    \caption{The reconstructed mass distribution 
for the $\kappa_0 = 2$ case.  
The reconstruction works well outside the critical region.  The
``reconstructed'' mass inside the critical lines shows up as noise,
since the method fails there. \label{fig:critical}}
  \end{minipage}
\end{figure*}

\section{The Lens Mapping Algorithm}\label{sec:algorithm}

In general the lens equation (\ref{eq:lens}) is a many-to-one mapping,
and consequently the inverse of the lens equation ${\mathbf x} = {\mathbf
x}({\mathbf y})$ has several branches, with no single branch
completely specifying the lens mapping. However, if the lens is
sub-critical, then the lens mapping is one-to-one everywhere.  In this
case both the mapping and its inverse are uniquely defined and
completely specify the mass distribution.  Here we show that in this
case it is possible to obtain the lens mapping completely from the
reduced shear, with the additional assumption that it goes to identity
sufficiently far away from the lens. 

For a non-critical lens, open curves in the source plane are mapped to
open curves in the lens plane.  In particular, any infinite straight
line is mapped to an infinite open curve in the lens plane.  In
Fig.~\ref{fig:grid}a, we schematically show the images of the two
perpendicular straight lines passing through the source at
$(y_1,y_2)$. The point of intersection of the two curves gives the
position of the image ($\bchi_1$, $\bchi_2$).

From (\ref{eq:matobs}) we can obtain the equation which the image
of an arbitrary infinite straight line in the source plane
satisfies in the lens plane. In particular, we obtain the equations
which map the coordinate grid lines of a Cartesian coordinate system
in the source plane to the lens plane
\begin{eqnarray}
dy_1 = 0 \mapsto \frac{dx_1}{dx_2} = \frac{\et}{1-\eo}\,\,; \quad dy_2 = 0 \mapsto \frac{dx_2}{dx_1} = \frac{\et}{1+\eo} \,\,.
\label{eq:vectorfield}
\end{eqnarray}
These are first-order ordinary differential equations and can be
uniquely integrated through any point in the lens plane. If we
consider the source position as the intersection of the lines $y_1=
{\rm constant}$ and $y_2={\rm constant}$ in the source plane, then the
image of this point will be the intersection of the images of these
two curves in the lens plane. In this manner we obtain a one-to-one
mapping from the lens to the source plane. Away from the lens
singularities we can integrate the equations (\ref{eq:vectorfield}) to
obtain the numerical solutions
\begin{eqnarray}
x_1 = X(x_2;{\bchi})\,\,; \quad x_2 = Y(x_1;{\bchi})
\label{eq:chis}
\end{eqnarray}
where $ X\,(x_2;\bchi)$ is the mapped curve for $dy_1=0$, passing
through the point $\bchi$ in the lens plane, and $ Y\,(x_1;\bchi)$ is
the mapped curve for $dy_2=0$, passing through the point $\bchi$ in the
lens plane (see Fig.~\ref{fig:grid}a).  The auxiliary variable
$\bchi$ in the above equations also explicitly represents the family
of curves which the equations (\ref{eq:vectorfield}) generate when
integrated with the initial conditions $ X\,(\chi_2) = \chi_1$ and
$Y\,(\chi_1) = \chi_2$, respectively, for the two equations.

In Fig.~\ref{fig:grid}b we show the image of a Cartesian grid that is
equally spaced in the source plane, lensed by a two-component model,
where both clusters are represented by Pseudo Isothermal Circular Mass
Distributions (PICMD) with the same core radius $r_c$, separated by
$10\,r_c$ along the $x_1$ axis. The central values of the dimensionless
surface density for the two components are $\kappa_0=0.8$ and $\kappa_0
=0.5$ respectively. Fig.~\ref{fig:grid}b illustrates the integral curves of
(\ref{eq:vectorfield}).

To obtain the source position corresponding to the image position
$\bchi$ we use the assumption that far away from the lens the lens
mapping is $ y_i \cong x_i$ (see Fig.~\ref{fig:grid}a). Therefore the
integrated curves should go to the original unperturbed source grid
lines at large $|{\mathbf x}|$.  This is ensured by the following
asymptotic conditions
\begin{eqnarray}
\label{eq:initialcondition1}
\lim_{x_2 \rightarrow \pm \infty} X(x_2; \bchi) = y_1 \,\,; \quad \lim_{x_1 \rightarrow \pm \infty} Y(x_1; \bchi) = y_2\,\,.
\end{eqnarray}
The lens equation can now be formally written as
\begin{eqnarray}
y_1 = X(\infty ;{\mathbf x})\,\,; 
\quad y_2 = Y(\infty ;{\mathbf x})\,\,,
\label{eq:asymptoticlenseqn}
\end{eqnarray}
where the auxiliary variables $\bchi$ are now
interpreted as the image coordinates ${\mathbf x}$.
The trace of the magnification matrix which is given in terms of
$\kappa$ as
\be
\frac{
\partial X(\infty ; {\bf x})}{\partial x_1} + 
\frac{\partial Y(\infty ; {\bf x})}{\partial x_2} = 2\,
   \left[1\,-\,\kappa({\mathbf x})\right]\,\,,  
\label{eq:kappa}
\en 
provides an estimate of $\kappa$ at any point.
Fig.~\ref{fig:contours} shows the lens mass
distribution thus reconstructed 
from the shear represented by Fig.~\ref{fig:grid}b,
assuming that the shear is noise-free and is known everywhere.

It might appear surprising that we have been able to obtain the lens
mapping solely from shear. To understand how, let us recall that our
assumption that the lens mapping becomes identity far away from the
lens removes the mass-sheet degeneracy We also note that it is
essentially equivalent to the assumption in Kaiser's method that the
mass density vanishes {\emph far away} from the center, though
quantitatively the notion of ``far'' is different in the two
methods. For a finite field we will see that the degeneracy reappears.

The above discussion applies to points away from the critical lines
of the lens mapping.
At these singularities, the eqs.~(\ref{eq:vectorfield})
become non-integrable. This becomes apparent
by writing the eqs.~(\ref{eq:vectorfield}) in
terms of $\gamma_i$s and $\kappa$,
\begin{eqnarray}
\frac{dx_1}{dx_2} = \frac{\gamma_2}{1-\kappa - \gamma_1}\,\,; \quad
\frac{dx_2}{dx_1} = \frac{\gamma_2}{1-\kappa + \gamma_1}\,\,.
\label{eq:field}
\end{eqnarray}
For a sufficiently smooth mass distribution, the potential and
consequently $\gamma_2$ is non-singular everywhere.  Since the
Jacobian of the lens mapping, $(1-\kappa)^2 - \gamma^2$, is positive
far away from the lens, we conclude that $(1-\kappa)^2 -\gamma_1^2 >
\gamma_2^2$.  From this inequality it is clear that $(1-\kappa)^2 -
\gamma_1^2$ can become zero only when the Jacobian $(1-\kappa)^2 -
\gamma^2$ also becomes zero.  Since $(1-\kappa)^2 - \gamma_1^2 =0$
implies that either $1-\kappa - \gamma_1 =0$ or $1-\kappa + \gamma_1
=0$, we conclude that both the equations (\ref{eq:vectorfield}) cannot
be integrated simultaneously, which implies that a correspondence
between the lens plane and source plane cannot be obtained.
Thus the LM algorithm cannot be applied in regions very close
to the critical lines of the Lens mapping.

\subsection{Finite Field, Critical Lens}\label{sec:finite}
The algorithm as described in the previous section is applicable to
cases where the reduced shear is known everywhere.  In practice,
this information is available only in a limited region of the lens
plane.  We show here that the LM algorithm can be generalized to this
case.
\label{eq:initialcondition}
For a finite field, eqs.~(\ref{eq:initialcondition1})
can be applied
only up to the edge of region in which data is available, and the
correct limit cannot be evaluated.  It is easy to see that
due to the unknown magnification factor at the edge, these equations
provide an incorrect measure of the derivatives in eq.~(\ref{eq:kappa}). 
 
Let us consider the differentials of $X$ and $Y$
(as defined in eq.~\ref{eq:chis}) at the edge ${\bf x}^{\rm e}$,
\begin{eqnarray}
\Delta X(x_2^{\rm e}, \bchi)= \frac{\Delta X}{\Delta
y_1}\Delta y_1 \,\,; \quad 
\Delta Y(x_1^{\rm e}, \bchi)= \frac{\Delta Y}{\Delta
y_2}\Delta y_2.
\label{eqn:diff}
\end{eqnarray}
Since at the edge of the image
the derivatives of $X$ and $Y$ are $>1$ we obtain
slightly higher values of $\Delta y_i$. Substituting the value of the
derivatives in eq.~(\ref{eqn:diff}) we obtain
\begin{eqnarray}
\Delta X(x_2^{\rm e}, \bchi) &=& [1 -\kappa\,({\bf x}^{\rm e})]\,
   [1-g_1\,({\bf x}^{\rm e})] \>\Delta y_1\,\,,\\
 \Delta Y(x_1^{\rm e}, \bchi) &=&
      [1-\kappa\,({\bf x}^{\rm e})]\,
   [1 + g_1\,({\bf x}^{\rm e})] \>\Delta y_2.
\end{eqnarray}
This gives us a way of obtaining the exact value of the derivatives of
the lens mapping, and thus completely removing the mass-sheet
degeneracy.  However, in reality we only know the reduced shear, and
consequently the correction factors at the edges, only up to a
factor of $1-\kappa$. For finite fields we can impose the boundary
condition $\kappa=0$ to obtain the mass distribution. This corrects
the generalized degeneracy up to a factor of $(1-\kappa)$.

This algorithm can be applied to a critical lens as well.  Since the
algorithm depends on our ability to integrate equations
(\ref{eq:vectorfield}) without hitting a singularity, we see that the
method can be used for the region which lie outside the critical
lines. In Fig.~\ref{fig:critical}, where the model has $\kappa_0=2$,
we see that our algorithm works well outside the critical region.  

\begin{figure*}
  \setlength{\figwidth}{\textwidth}
  \addtolength{\figwidth}{-\columnsep}
  \setlength{\figwidth}{0.5\figwidth}
  
  \begin{minipage}[t]{\figwidth}
    \mbox{}
    \vskip -7pt
    \centerline{\includegraphics[width=0.88\linewidth,angle=0]{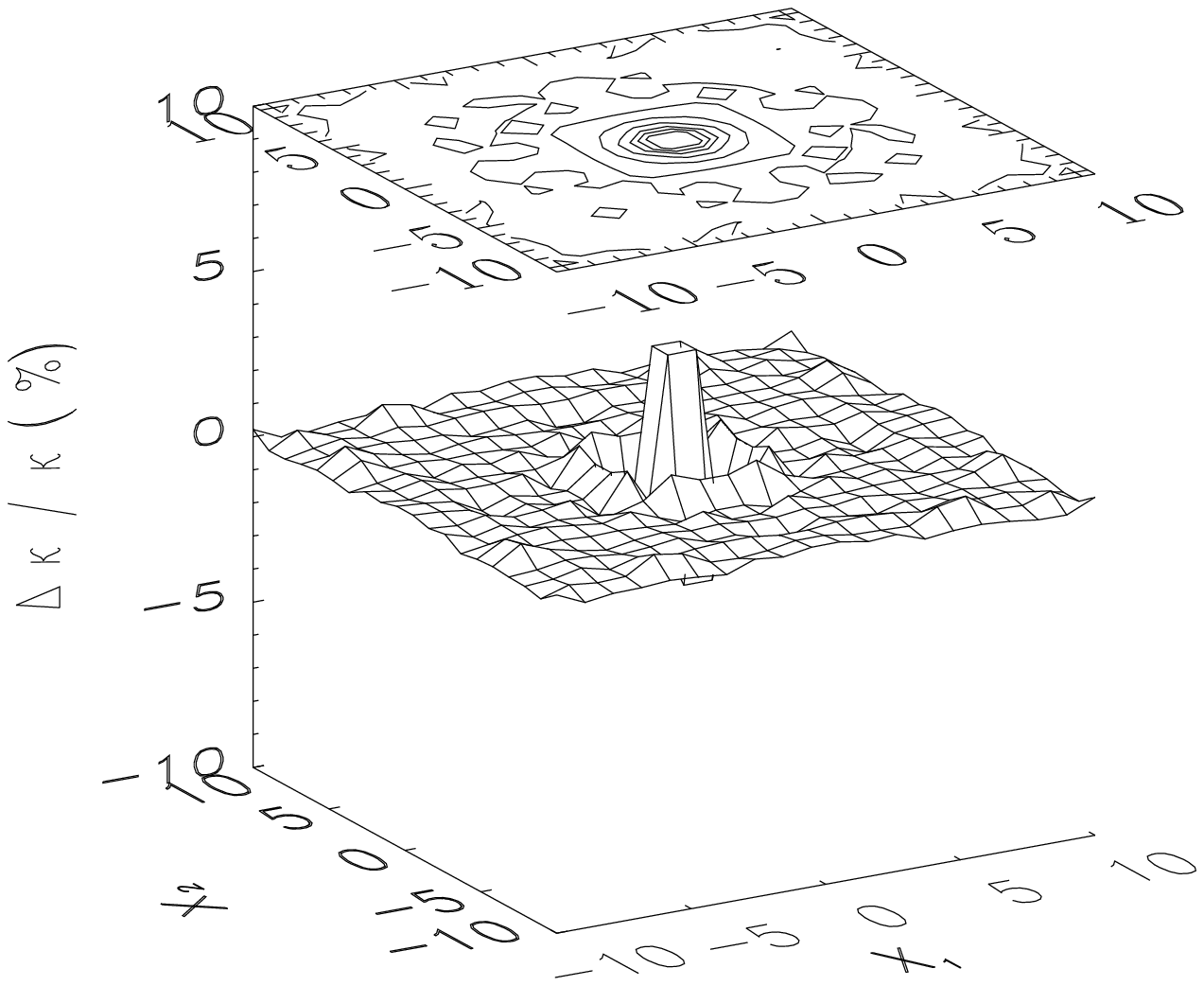}}
\caption{The effect of the finite sampling of shear: the
percentage error in the reconstructed mass distribution (using the LM
algorithm), in the case where the reduced shear is known on a grid of
size $20\arcsec\times 20 \arcsec$, and there is no noise.  The mass is
reconstructed using the LM algorithm on a $20\times 20$ grid.  The
excess noise near the center of the distribution is due to the fact
that the reduced shear is most rapidly varying close to the center,
where the interpolated shear values are consequently noisier.  Note
that the rms noise outside the central region is less than 1\%.
\label{fig:errorgrid}}
  \end{minipage}
  \hfill
  \begin{minipage}[t]{\figwidth}
    \mbox{}
    \vskip -7pt
    \centerline{\includegraphics[width=0.88\linewidth,angle=0]{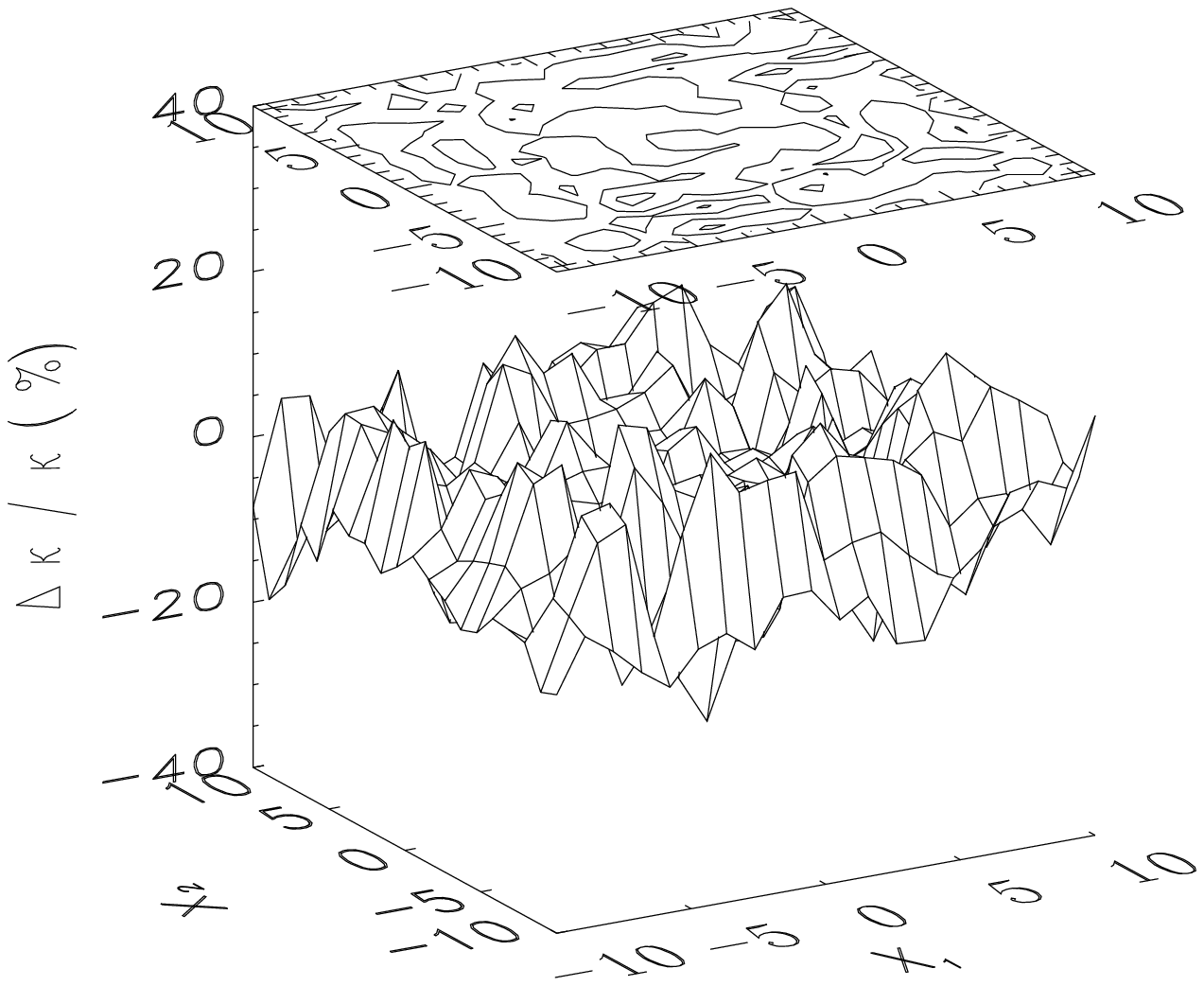}}
    \caption{
The effect of the finite sampling and noise: the
percentage error in the reconstructed mass distribution in the same
case as the last one, but with noise with a constant signal-to-noise
ratio of $10$ added to the shear field before the application of
interpolation.  \label{fig:errornoise}
}
  \end{minipage}
\end{figure*}

\section{Noisy and Discretely Sampled Data}\label{sec:noisy}

We now present a few examples to illustrate the application of the LM
algorithm to noisy data.  Since we wish to compare the original mass
distribution with that reconstructed by the algorithm, we correct for
the mass-sheet degeneracy in all the examples by the prescription
given in \S ~\ref{sec:finite}. We consider a single component lens
modeled as PICMD. We simulate the signal by analytically calculating
the reduced shear from a model. We illustrate the effect of finite
sampling of data and the presence of noise with separate examples.

We first consider the case where the reduced shear is known at all the
points from an analytical model.  We choose a lens with the central,
dimensionless surface mass density given by $\kappa_0=0.7$, and
reconstruct the mass distribution by applying the LM algorithm on a
$20 \times 20$ grid.  The noise in the reconstruction, which is very
small everywhere ($\simlt 1\%$), can be entirely accounted for as
arising from the inaccuracies in the evaluation of various integrals
and derivatives required by the algorithm.

To illustrate the effect of finite sampling of shear in the image
plane, we consider a model with a lens velocity dispersion of $\sigma_v =
1100$ km/s and with a core radius $R_c=50$ kpc.  The source galaxies
are assumed to be at the redshift $z_s = 1$ and the lens is at 
$z_l=0.2$.  We sample
the reduced shear in pixels of size $20^{\prime\prime} \times
20^{\prime\prime}$. Using a bicubic spline interpolation routine to
evaluate the reduced shear at all the points, the mass distribution is
reconstructed using the LM algorithm on a $20 \times 20$ grid as
before.  The resulting percentage error 
in the reconstructed mass distribution (given by $100
\times(\kappa_{\rm rec} - \kappa_{\rm true})/\kappa_{\rm true}$) is shown in
Fig.~\ref{fig:errorgrid}.  We note that the excess noise near the
center of the distribution is mainly due to the fact that the reduced
shear is most rapidly varying close to the center and therefore the
interpolated shear values are noisier at those points. This noise will
vanish if the grid on which the shear is evaluated were made finer,
but then in the real world, one is limited by the number of galaxies
over which the value of the reduced shear has to be averaged, and we
cannot use a substantially finer grid without increasing shot noise.

As our last example we consider the same case as the last one, but add
a uniform noise with a constant signal-to-noise ratio of $10$ to the
given shear field, before the application of interpolation.  The noise
in the reconstructed mass distribution (Fig.~\ref{fig:errornoise}) can
be seen to be around twice that of the noise in the shear field.

\section{Conclusions}

We have described a simple algorithm for mass reconstruction of a
cluster lens by directly obtaining the lens mapping from the measured
reduced shear and taking its derivatives to compute the surface mass
density $\kappa$. 

The method works best for sub-critical lenses, where it can give the
mass distribution at all points, but it can work for critical lenses
as well in limited regions of the lens plane, away from the critical
lines.  The algorithm is shown to have a mass-sheet degeneracy of the
same type as exists in the other methods of reconstruction if the
field in which the reduced shear is available is finite.  We have
tested the algorithm on discretely sampled noisy (simulated) data and
have found that it reproduces the mass distribution within acceptable
limits.

\acknowledgements 
TDS thanks the University Grants Commission (India) and IUCAA for
providing support for this work. We thank Yuri A. Shchekinov and
Prof. J. Ehlers for useful discussions, and an anonymous referee
for helpful suggestions.

\end{document}